\documentclass[journal,twoside,web]{ieeecolor}
\usepackage{etoolbox}
\makeatletter
\@ifundefined{color@begingroup}%
{\let\color@begingroup\relax
\let\color@endgroup\relax}{}%
\def\fix@ieeecolor@hbox#1{%
\hbox{\color@begingroup#1\color@endgroup}}
\patchcmd\@makecaption{\hbox}{\fix@ieeecolor@hbox}{}{\FAILED}
\patchcmd\@makecaption{\hbox}{\fix@ieeecolor@hbox}{}{\FAILED}

\usepackage{tmi}
\usepackage{amsmath,amssymb,amsfonts}
\usepackage{algorithmic}
\usepackage{cite}
\usepackage{graphicx}
\usepackage{multirow}
\usepackage{ragged2e}
\usepackage[linesnumbered,ruled,vlined]{algorithm2e}
\usepackage{url} 

\def\BibTeX{{\rm B\kern-.05em{\sc i\kern-.025em b}\kern-.08em
    T\kern-.1667em\lower.7ex\hbox{E}\kern-.125emX}}
\begin{document}
\title{Interactive Gadolinium-Free MRI Synthesis: A Transformer with Localization Prompt Learning}
\author{Linhao~Li,~Changhui~Su,~Yu~Guo,~Huimao~Zhang,~Dong~Liang,~\IEEEmembership{Senior Member, IEEE}~and~Kun Shang
\thanks{This study was supported in part by the Natural Science Foundation of China under grants No.62125111, No.62331028, No.82430062.}
\thanks{Linhao~Li,~Changhui~Su~and~Yu~Guo contributed equally to this work.}
\thanks{Corresponding authors are~Dong~Liang~and~Kun~Shang. E-mail: dong.liang@siat.ac.cn; kunzzz.shang@gmail.com}
\thanks{Linhao~Li and Changhui~Su are with the School of Artificial Intelligence, Hebei University of Technology, Tianjin 300401, China.}
\thanks{Yu~guo and~Huimao~Zhang are with the Department of Radiology, The First Hospital of Jilin University, Changchun, Jilin, China.}
\thanks{Changhui~Su,~Liang~Dong and Kun Shang are with the Research Center for Medical AI, Shenzhen Institute of Advanced Technology, Chinese Academy of Sciences, Shenzhen 518055, China.}
}

\maketitle

\begin{abstract}
Contrast-enhanced magnetic resonance imaging (CE-MRI) is crucial for tumor detection and diagnosis, but the use of gadolinium-based contrast agents (GBCAs) in clinical settings raises safety concerns due to potential health risks. To circumvent these issues while preserving diagnostic accuracy, we propose a novel Transformer with Localization Prompts (TLP) framework for synthesizing CE-MRI from non-contrast MR images. Our architecture introduces three key innovations: a hierarchical backbone that uses efficient Transformer to process multi-scale features; a multi-stage fusion system consisting of Local and Global Fusion modules that hierarchically integrate complementary information via spatial attention operations and cross-attention mechanisms, respectively; and a Fuzzy Prompt Generation (FPG) module that enhances the TLP model's generalization by emulating radiologists' manual annotation through stochastic feature perturbation. The framework uniquely enables interactive clinical integration by allowing radiologists to input diagnostic prompts during inference, synergizing artificial intelligence with medical expertise. This research establishes a new paradigm for contrast-free MRI synthesis while addressing critical clinical needs for safer diagnostic procedures. Codes are available at https://github.com/ChanghuiSu/TLP.
\end{abstract}


\section{Introduction}
\label{sec:introduction}
\IEEEPARstart{C}{ontrast-enhanced} magnetic resonance imaging (CE-MRI) is a powerful diagnostic tool widely used in oncology, neurology, and vascular imaging, where accurate lesion delineation is essential for effective treatment planning and monitoring~\cite{grobner2006gadolinium,warntjes2018synthesizing}. This technique involves injecting specialized contrast agents, typically gadolinium-based contrast agents (GBCAs), into the bloodstream to enhance the visualization of specific tissues or blood vessels~\cite{zahra2007dynamic}. By significantly improving the contrast between normal and abnormal tissues, CE-MRI facilitates the precise detection of tumors, inflammation, and other pathological conditions~\cite{ silver1997sensitivity, khan2014molecular, moya2024exogenous}. 

Although GBCAs used in CE-MRI have significant clinical value, there is public concern about their safety, particularly their association with nephrogenic systemic fibrosis and the potential for gadolinium retention in the body~\cite{thomsen2006nephrogenic,marckmann2006nephrogenic,forghani2016adverse,olchowy2017presence,semelka2016gadolinium,moya2024exogenous}. Moreover, compared to non-contrast-enhanced MRI, the use of GBCAs significantly extends the scanning time and demands higher technical skills from the operators~\cite{purysko2011focal}. This also means that CE-MRI is not always available, and due to the high cost of data acquisition, re-scanning to recover missing sequences is not usually an economical option~\cite{zhang2024unified}. Therefore, reducing dependence on GBCAs and addressing the issue of data loss in CE-MRI have become a research area of great interest.

Recently, efforts have been made to synthesize contrast-enhanced images from non-contrast images to reduce the use of GBCA in CE-MRI \cite{jiao2023contrast,ye2013modality,huang2017simultaneous}. These methods leverage deep learning technologies to learn the complex nonlinear mappings between contrast-enhanced and non-contrast images \cite{van2015cross,zhang2023synthesis,zhao2020tripartite}. By training on large datasets of paired CE-MRI and non-contrast images, these models can predict the appearance of contrast-enhanced tissues in non-contrast images, thereby simulating the effects of GBCAs without actual injection. One of the key advantages of these synthetic approaches is their potential to reduce the risks associated with GBCAs while minimizing scanning time and technical demands, thereby making CE-MRI more accessible and cost-effective~\cite{li2022virtual}.

Despite these remarkable advancements, CE-MRI synthesis remains a significant challenge. The often imperceptible structural distinctions between pathological lesions and healthy tissues significantly impede the achievement of accurate image synthesis.  And the intricate nonlinear relationship between contrast-enhanced and non-contrast images frequently results in diminished detail preservation and compromised accuracy in synthesized outputs, especially within tumor regions~\cite{zhao2020tripartite,sevetlidis2016whole,chartsias2017multimodal,joyce2017robust}.  Synthesized images demonstrate strong temporal consistency with actual CE-MRI data, they tend to underestimate tumor volumes~\cite{ji2022synthetic}. Moreover, the presence of false-positive enhancements or distortions in tumor margins during detection processes exacerbates the potential for clinical misdiagnosis~\cite{preetha2021deep}. Hence, \textbf{\textit{to improve the accurate expression of tumors in the synthetic CE-MRI images is the key to this problem.}}

In this paper, we propose the Transformer with Localization Prompt (TLP) framework for synthesizing CE-MRI from non-contrast MRI. TLP implements a spatially aware prompt mechanism that encodes tumor location information, enabling more accurate synthesis of lesion areas and resolving the common issue of tumor boundary distortion. This study is of significant importance for eliminating the risks associated with GBCAs and shortening scan times. 

The main contributions of this paper are summarized as follows:

\begin{itemize}    
    \item{The TLP framework is interactive, enabling human prior knowledge to assist in the generation process. This interactivity unlocks a range of exciting applications, enhancing not only the quality of generated results through prompts but also providing improved support for downstream tasks.}
    
    \item{We design a Fuzzy Prompt Generation (FPG) module to simulate radiologists' preliminary assessments of non-contrast images (T1-weighted, T2-weighted), automatically generating initial localization of suspected lesion areas. FPG also prevents the model from relying on fixed prompts, thus enhancing the model’s generalization ability.}

    \item{We propose a novel Transformer-based backbone that introduces a multi-scale feature processing mechanism to fully leverage the representational capacity of spatial dependencies in medical imaging data. This approach demonstrates significant advantages in handling large and complex tumors.}
    
    \item{We develop the Local Fusion (LF) and Global Fusion (GF) modules, effectively utilizing complementary information from different modalities. These modules are designed to synergistically integrate fine-grained local details with comprehensive contextual information, demonstrating exceptional performance in handling lesions of varying sizes.}
\end{itemize}

\section{RELATED WORK}

\subsection{Convolutional models}
In recent years, CNN-based methods~\cite{isola2017image,zhu2017unpaired} have driven significant advancements in medical image translation. Notable contributions include Zhou et al.'s HiNet~\cite{zhou2020hi}, which employs a convolution-based encoder and fusion network to generate target modalities from dual input sources. Building upon this foundation, Li et al.'s MMgSN-Net~\cite{li2022virtual} and Zhang et al.'s unified framework~\cite{zhang2024unified} have further enhanced the convolutional backbone architecture by integrating self-attention mechanisms for intermediate feature modulation. However, despite their demonstrated efficacy, CNN-based approaches exhibit several inherent limitations that warrant consideration.

A primary constraint lies in the limited generalization capability of convolution operations when processing non-standard anatomical structures, often resulting in suboptimal representation of irregular or atypical features. Furthermore, the intrinsic local receptive field of CNNs imposes fundamental restrictions on capturing long-range spatial dependencies, which are particularly crucial in medical imaging where global context and distant structural relationships frequently determine diagnostic accuracy~\cite{wang2018non,kodali2017convergence}. While the incorporation of self-attention modules has partially mitigated these limitations, empirical evidence suggests that their effectiveness remains constrained due to the convolutional generation of low-level feature maps, ultimately restricting global context sensitivity~\cite{atli2024i2i,xie2021cotr,chen2021transunet}.

\subsection{Transformer models}
To improve context representation in medical image analysis, Transformer-based architectures have gained prominence due to their exceptional ability to capture long-range dependencies. However, initial versions of these models faced substantial computational limitations, restricting their effectiveness. For example, ResViT, developed by Dalmaz et al.~\cite{dalmaz2022resvit}, incorporates two Transformer modules with shared parameters into the convolutional generator of CycleGAN. This design requires additional downsampling before each Transformer module to manage the quadratic computational complexity tied to feature size, though this process risks losing vital information through excessive feature compression.

To overcome these computational challenges, several optimized Transformer models have emerged. The Swin Transformer, introduced by Liu et al.~\cite{liu2021swin}, utilizes a sliding window mechanism that limits self-attention computations to non-overlapping windows, markedly enhancing computational efficiency. While such innovations provide useful inductive biases, they may compromise the Transformer’s capacity to capture broad contextual relationships. Similarly, 
Zamir et al.~\cite{zamir2022restormer} propose Restormer, which reduces computational complexity by refining multi-head attention and feedforward networks while preserving long-range pixel modeling for large-scale image restoration.
These developments highlight the possibility of balancing computational efficiency with the retention of Transformers’ strengths in medical image analysis.

\subsection{Other Deep Generative models}
Recent research endeavors have explored the integration of advanced generative paradigms, particularly diffusion models~\cite{ozbey2023unsupervised,jiang2024fast} and reinforcement learning approaches~\cite{xu2021synthesis}, into MRI synthesis tasks. While these methods demonstrate promising capabilities in generating high-quality medical images, they present substantial computational challenges that limit their practical applicability. The implementation of diffusion models, in particular, necessitates extensive computational resources, with both training and inference processes requiring orders of magnitude more time compared to conventional methods. Specifically, for equivalent image generation tasks, diffusion-based approaches typically exhibit computational times that are several hundredfold greater than those of alternative methodologies, imposing significant constraints on their clinical deployment and scalability.

\begin{figure*}[!t] 
\centering
\includegraphics[width=\textwidth]{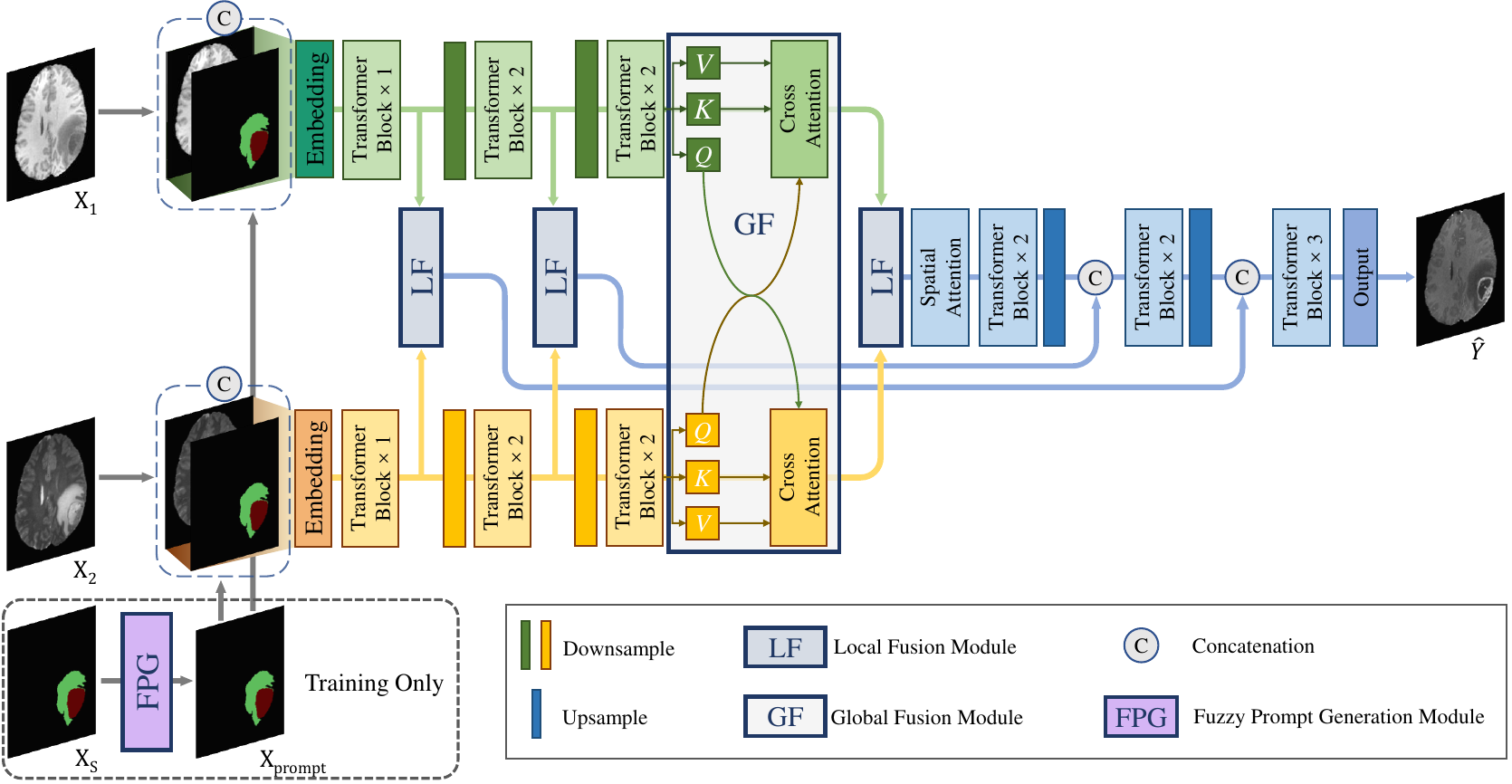}
\caption{Schematic diagram of the proposed TLP for MRI modality synthesis, taking dual-modality input as an example. The model is built on a multi-stage Transformer architecture, establishing a symmetric encoder-decoder framework that processes features at various resolutions. Central to its design is the interaction between local and global features, facilitated by the Local Fusion (LF) and Global Fusion (GF) modules. These modules integrate information from different modalities, capturing both local and global contexts. During the training phase, a prompt generation module is employed to generate tumor location prompts, guiding feature generation and enhancing model interactivity.}
\label{fig: framework}
\end{figure*}

\subsection{Prompt Learning}
Prompt learning, originally developed in the domain of natural language processing (NLP), has undergone significant expansion into visual generation tasks through advancements in multimodal learning frameworks such as CLIP~\cite{radford2021learning} and Stable Diffusion~\cite{rombach2022high}. This paradigm shift has enabled sophisticated text-to-image generation systems that utilize natural language descriptions as semantic prompts to guide image synthesis, as demonstrated in notable works like hierarchical text-conditional image generation~\cite{ramesh2022hierarchical} and photorealistic text-to-image diffusion models~\cite{saharia2022photorealistic}.

Parallel to textual prompts, visual conditioning mechanisms have emerged as powerful tools for content control, employing various modalities including segmentation masks, key points, and sketches. A seminal example is SPADE~\cite{park2019semantic}, which leverages semantic segmentation maps as conditional inputs to generate high-fidelity scene images. The field has further evolved with multimodal approaches like GLIDE~\cite{nichol2021glide}, which integrates CLIP text encoders with segmentation masks to achieve comprehensive prompt-driven image generation.

Despite these advancements, the medical imaging domain presents unique challenges for prompt-based generative models. The limited availability of large-scale multimodal datasets has hindered the development of effective text-controlled generation systems in this specialized field, creating a significant gap between general computer vision applications and medical image synthesis capabilities.

The method proposed in this paper draws inspiration from existing prompt learning paradigms, emphasizing the use of regions of interest (ROIs) as structured visual prompts. Unlike traditional text-based prompts, ROIs introduce additional spatial semantic constraints—such as lesion location and shape—allowing for more precise guidance of the generative model toward lesion areas. This design seamlessly integrates the lightweight conditional control benefits of prompt learning with the critical need for spatial consistency in image generation tasks, providing an efficient and effective solution for enhancing the synthesis of magnetic resonance imaging.

\section{PROPOSED METHOD}
This section presents the proposed TLP for MRI modality synthesis tasks. As illustrated in Fig.~\ref{fig: framework}, the architecture of TLP comprises a Transformer-based backbone\footnote{The Transformer module adopts the design of Restormer~\cite{zamir2022restormer}, a simplified Transformer that reduces the computational cost, enabling its use across multiple levels.} integrated with three specialized submodules: (1) Fuzzy Prompt Generation (FPG) module generates adaptive prompt information during the training phase; (2) Local Fusion (LF) Module effectively integrates local-level features; and (3) Global Fusion (GF) Module facilitates comprehensive global information exchange. 
The following subsections provide detailed descriptions of each module's architecture and functionality.

\subsection{Fuzzy Prompt Generation Module}
The Fuzzy Prompt Generation (FPG) module is designed to simulate radiologists’ preliminary assessments of non-contrast images (T1-weighted, T2-weighted) by performing initial localization of suspected lesion regions within these scans. This component operates exclusively during the training phase, generating blurred region-of-interest (ROI) prompts through its convolution-based stochastic scaling algorithm, which is detailed in Algorithm \ref{alg:crs}. It defines a series of convolution kernels to process the input in different directions:
\begin{equation}
\mathcal{K} = \{ {E}_{2,2} + \mathbf{E}_{i,j} \mid (i,j) \in S \} \cup \{ \sum_{(i,j) \in T} \mathbf{E}_{i,j}\}, 
\end{equation}
where $S = \{1,2,3\}^2 \setminus \{(2,2)\}$, $T=\{1,2,3\}^2\setminus \{1,3\}^2$, and 
\(\mathbf{E}_{i,j}\) represents a $3 \times 3$ basic matrix where the value is 1 at position \((i,j)\) and 0 at all other positions.
\begin{algorithm}
\SetAlgoLined 
\caption{Convolution-based random scaling algorithm}
\label{alg:crs}
\KwIn{Input label $X_{S} \in \mathbb{R}^{m \times n}$, the probability of dilation $p$, the probability of losing a prompt $q$, scaling times $t \in \mathbb{Z}^{n}$, convolution kernels $\mathcal{K}$}
        
\KwOut{The generated prompt $X_{\text{prompt}}$}
\SetAlgoLined 
$q'  \gets  \text{rand}(0, 1)$\\
\uIf{$q' < q$}{
    $X_{\text{prompt}} \gets \text{zeros}(m,n)$
}\Else{
    \For{$i \in [0, t) $}{
        $p' \gets \text{rand}(0, 1)$\\
        $K \gets \text{randomly select a kernel from}~\mathcal{K}$\\
        \uIf{$p' < p$}{
            $X_{S} \gets \text{double}(K \ast X_{S}> 0)$ 
        }\Else{
            $X_{S} \gets \text{double}(K \ast X_{S}> 1)$ 
        }
    }
$X_{\text{prompt}} \gets X_{S}$
}
\end{algorithm}

By configuring the dilation probability $p$ and the number of dilation iterations $t$, we simulate the suspicious ROIs manually delineated by physicians (such as the suspected lesion regions $X_S$ in non-contrast images T1-weighted ($X_1$) and T2-weighted ($X_2$) in Fig.~\ref{fig: framework}). A random omission probability $q$ is introduced to stochastically exclude these hints, preventing the model from over-relying on the prompts and thereby enhancing its robustness. During the training phase, we set $q=0.5, p=0.9, t=5$ to ensure that the generated ROIs 
from FPG are visually consistent with manually annotated results.

\subsection{Local Fusion Module}
The Local Fusion (LF) Module is a multi-modal feature integration framework designed to effectively combine Transformer-based features from heterogeneous modalities through a series of hierarchical operations. It incorporates convolution, concatenation, and element-wise maximum operations to fuse features. The mathematical representation of this process is as follows:
\begin{equation}\label{localfusion}
F_{LF}^l = \mathbf{C}(concat(F_{1}^l,F_{2}^l, max(F_{1}^l,F_{2}^l))),
\end{equation}
where $F_{i}^l$ means the specific $l$-th Transformer-based features from the $i$-th modality, $max$ refers to the element-wise maximum operation applied across the fused features, $concat$ denotes concatenation, and $C$ signifies the convolution operation. The LF module leverages the sensitivity of convolutions to local details, performing an initial fusion of shallow features from different modalities. This fusion helps the model to effectively utilize complementary information from each modality, enhancing its ability to capture local patterns and fine-grained features.

\subsection{Global Fusion Module}
The Global Fusion (GF) module is designed to integrate global information between features from two different input modalities by employing a cross-attention mechanism. Concretely, the final Transformer-based features from each modality, $F_i^T, i=1, 2$, are transformed into Query (Q), Key (K), and Value (V) components using two separate projection functions $\mathcal{P}_{qkv}^{i}$:
\begin{equation}
\{ Q_{i}, K_{i}, V_{i} \} = \mathcal{P}_{qkv}^{j}(F_{i}^T).
\label{gfm}
\end{equation}

This step is crucial as it allows the features to be converted into components suitable for attention computation.
Then, the queries $Q_1$ and $Q_2$ from the two modalities are exchanged to facilitate interaction:
\begin{equation}
F_{2}^G = \text{softmax}\left(\frac{Q_{1} K_{2}^\top}{\sqrt{d_k}}\right)V_{2},
\label{gfm3}
\end{equation}
and
\begin{equation}
F_{1}^G = \text{softmax}\left(\frac{Q_{2} K_{1}^\top}{\sqrt{d_k}}\right)V_{1},
\label{gfm4}
\end{equation}
where $d_k$ is the scaling factor. GF leverages cross-attention to enable global information integration between features from different modalities, enhancing the model’s ability to capture comprehensive and meaningful representations. The fused features are then further refined by LF to produce the final output.

\subsection{Loss Function}
In addition to using traditional pixel loss, we employ a discriminator $D$ based on PatchGAN~\cite{zhu2017unpaired} to distinguish between real enhanced contrast image $Y$ and synthetic image $\hat{Y}$.  Hence, the whole loss function is:
\begin{equation}
\begin{aligned}
\mathcal{L} & = \mathbb{E}[\| D(\hat{Y}) \|_2^2] + \mathbb{E}[\| D(Y) - 1 \|_2^2]\\
            & +\mathbb{E}[\| D(\hat{Y}) - 1 \|_2^2] + \lambda \mathbb{E}[\| \hat{Y} - Y \|_1].
\end{aligned}
\end{equation}
Where $\lambda$ is a hyperparameter used to balance the adversarial loss and the pixel loss.

\begin{figure*}[ht!]
\centering
\includegraphics[width=\textwidth]{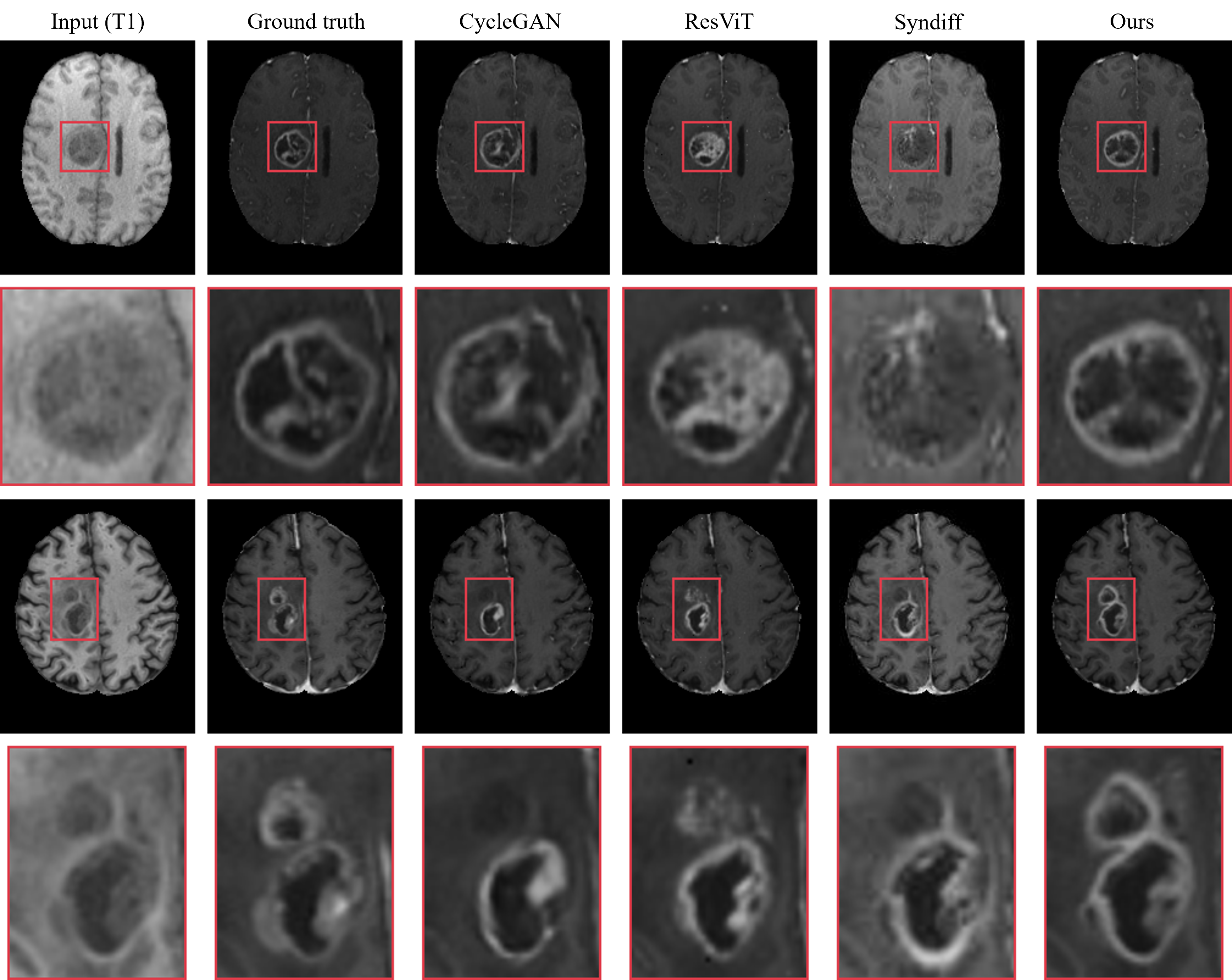}
\caption{Visual comparison of synthesis results from all one-to-one models on the BraTS 2021 dataset. Each model takes T1 as input simultaneously to synthesize T1ce.}
\label{fig:T1_T1ce}
\end{figure*}

\begin{figure*}[ht!]
\centering
\includegraphics[width=\textwidth]{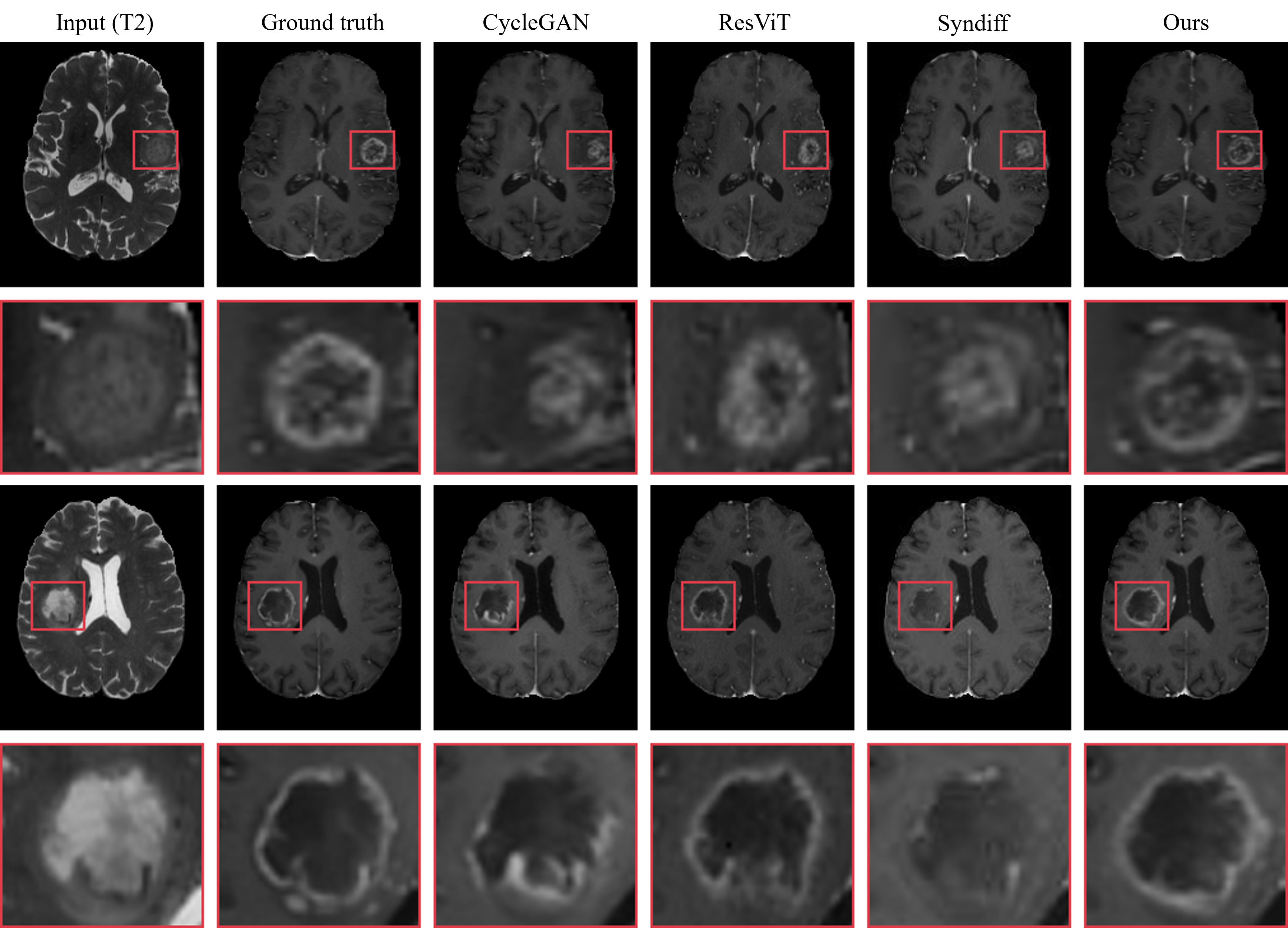}
\caption{Visual comparison of synthesis results from all one-to-one models on the BraTS 2021 dataset. Each model takes T2 as inputs simultaneously to synthesize T1ce.}
\label{fig:T2_T1ce}
\end{figure*}

\begin{figure*}[ht!]
\centering
\includegraphics[width=\textwidth]{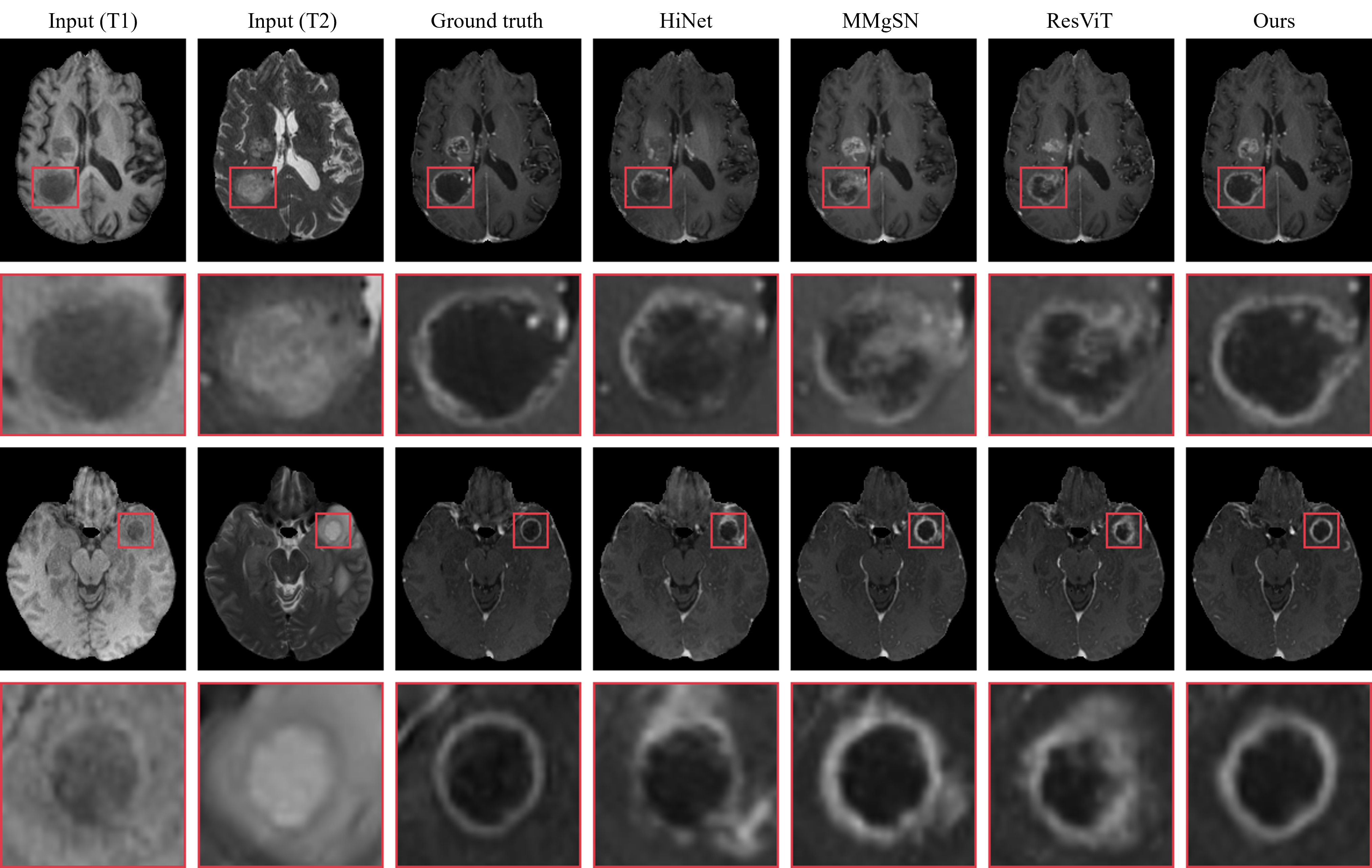}
\caption{Visual comparison of synthesis results from all many-to-one models on the BraTS 2021 dataset. Each model takes T1 and T2 as inputs simultaneously to synthesize T1ce.}
\label{fig:T1T2_T1ce}
\end{figure*}

\section{EXPERIMENTS AND RESULTS}
\subsection{Dataset}


The experiments were conducted on the dataset provided by the 2021 Multimodal Brain Tumor Segmentation Challenge (BraTS2021). The BraTS2021 dataset consists of multiparametric MRI (mpMRI) scans from 1,251 patients with diffuse glioma, including four modalities: T1-weighted (T1), contrast-enhanced (T1ce), T2-weighted (T2), and FLAIR. Due to variations in image quality, we selected 1,000 cases for our experiments, and the data list will be released alongside our code. All multimodal images in the BraTS dataset have been co-registered to a common anatomical template, resampled to a resolution of 1×1×1 mm³, and skull-stripped.

We used 2D axial slices of the volumes as input. Each 2D axial slice (240×240) was cropped to 192×160 from the center of the image. The original intensity values were linearly scaled to the range of [-1, 1]. Additionally, 850 cases were randomly assigned to the training set, 20 to the validation set, and 130 to the test set. For each case, 100 axial slices containing brain tissue were selected for evaluation.

\subsection{Competing Methods}
We conducted comprehensive comparisons between our proposed model and several state-of-the-art (SOTA) cross-modality synthesis approaches: CycleGAN~\cite{zhu2017unpaired}, HiNet~\cite{zhou2020hi}, MMgSN-Net~\cite{li2022virtual}, ResViT~\cite{dalmaz2022resvit}, and SynDiff~\cite{ozbey2023unsupervised}. All comparative methods were implemented using their official open-source codes with configurations strictly adhering to their original publications. 

\begin{table*}[ht!]
\centering
\caption{Comparison of Model Performance on Various Metrics}
\label{tab:results}
\begin{tabular}{l|lll|lll|lll}
\hline
\multicolumn{1}{c|}{\multirow{2}{*}{Method}} & \multicolumn{3}{c|}{T1→T1ce}                                                   & \multicolumn{3}{c|}{T2→T1ce}                                                   & \multicolumn{3}{c}{T1,T2→T1ce}                                                \\
\multicolumn{1}{c|}{}                        & \multicolumn{1}{c}{PSNR$\uparrow$} & \multicolumn{1}{c}{SSIM$\uparrow$} & \multicolumn{1}{c|}{NMSE$\downarrow$} & \multicolumn{1}{c}{PSNR$\uparrow$} & \multicolumn{1}{c}{SSIM$\uparrow$} & \multicolumn{1}{c|}{NMSE$\downarrow$} & \multicolumn{1}{c}{PSNR$\uparrow$} & \multicolumn{1}{c}{SSIM$\uparrow$} & \multicolumn{1}{c}{NMSE$\downarrow$} \\ \hline
HiNet       & -          & -       & -        & -       & -       & - & 26.09$\pm$3.18 & 0.843$\pm$0.057 & 0.663    \\ 
MMgSN-Net   & -    & -       & -      & -          & -       & -        & 29.36$\pm$3.50 & 0.891$\pm$0.047 & 0.263 \\ 
CycleGAN    & 25.76$\pm$3.64 & 0.842$\pm$0.064 & 0.430 & 26.13$\pm$3.21 & 0.836$\pm$0.058 & 0.301 & -       & -       & -    \\ 
SynDiff     & 21.75$\pm$4.15 & 0.801$\pm$0.072 & 0.998 & 24.03$\pm$4.01 & 0.825$\pm$0.060 & 0.609 & -       & -       & -    \\ 
ResViT      & 27.27$\pm$3.48 & 0.860$\pm$0.047 & 0.280 & 26.29$\pm$3.59 & 0.831$\pm$0.053 & 0.370 & 28.55$\pm$3.26 & 0.875$\pm$0.046 & 0.263 \\ 
Ours      & \textbf{28.81$\pm$3.25}      & \textbf{0.885$\pm$0.045}& \textbf{0.185} & \textbf{28.60$\pm$3.40}      & \textbf{0.880$\pm$0.050}& \textbf{0.186} & \textbf{30.22$\pm$3.45}      & \textbf{0.909$\pm$0.040}& \textbf{0.138} \\ \hline
\end{tabular}
\end{table*}

\subsection{Implementation Details}

All experiments were conducted on Nvidia RTX A6000 GPUs using the PyTorch framework, with the Adam optimizer \cite{kingma2014adam} configured as $\beta_1 = 0.5$ and $\beta_2 = 0.999$. To ensure convergence and optimal performance, we maintained the hyperparameter settings from the original publications for each method while implementing rigorous validation set monitoring throughout the training process.

The initial learning rate for our TLP model was set to 0.0001. All models were trained for 40 epochs with a fixed learning rate, followed by 40 additional epochs with linear learning rate decay to zero. Notably, ResViT followed the approach in the original paper, fine-tuning the Transformer module on ImageNet pre-trained weights. SynDiff was trained for 30 epochs, as the original literature suggests that its training duration is approximately half that of non-diffusion methods. Despite the shorter number of training epochs, SynDiff's training time still required about two weeks, significantly longer than the two to three days required for other comparison methods. Except for CycleGAN, the loss function weight $\lambda$ for all methods was uniformly set to 100. Convergence was observed on the validation set for all methods.

\subsection{Evaluation Metrics}
This study employs three widely-used metrics for quantitative evaluation of synthesis performance: 1)Mean Squared Error (MSE), 2) Peak Signal-to-Noise Ratio (PSNR), 3) Structural Similarity Index Measure (SSIM)~\cite{wang2004image}.

\subsection{Main Results}
Our experimental evaluation on the BraTS2021 dataset assessed three distinct cross-modality synthesis tasks: two single-input scenarios (T1 to T1ce and T2 to T1ce) and one multi-input scenario (T1 and T2 to T1ce). When used for single-input tasks, TLP activates only one input branch, and the two fusion modules are replaced with identity transformations.

\textbf{Qualitative performance:}
We demonstrate the visual comparisons between our method and comparative methods in Figures \ref{fig:T1_T1ce}, \ref{fig:T2_T1ce}, and \ref{fig:T1T2_T1ce}.

Fig.~\ref{fig:T1_T1ce} and Fig.~\ref{fig:T2_T1ce} show a comparison of the generated results between our method and CycleGAN, ResViT, and SynDiff in the \textbf{single-input scenario}. Clearly, our method synthesizes the most realistic details, with fewer artifacts and clearer tumor contours, which are critical for accurate diagnosis. While there is little difference in the healthy tissue synthesis across the different methods, our model outperforms the competing methods when it comes to lesion areas. It is noteworthy that during the testing phase, we did not use any prompts, yet our model still demonstrated strong attention to abnormal regions. Specifically, when only T1-weighted is used as input, as shown in the second-row example of Fig.~\ref{fig:T1_T1ce}, only our method does not miss small lesions. When T2-weighted is used as input, due to the more prominent lesions on the T2 image, as shown in Fig.~\ref{fig:T2_T1ce}, all methods do not miss any lesions. However, our model still achieves the most perfect detail synthesis and clear tumor contours, demonstrating significant advantages when handling lesions of different sizes.

Fig.~\ref{fig:T1T2_T1ce} compares our method with HiNet, MMgSN-Net, and ResViT in the \textbf{multi-input scenario}. While all methods benefit from additional input information, our model consistently produces the highest visual quality. Regardless of tumor size, our model can synthesize accurate tumor contours, while other methods generate discontinuous contours and incorrect enhancement regions. Notably, the Transformer-based ResViT model struggles with complex tumor structures due to excessive feature compression.

\textbf{Quantitative performance:}
As shown in Table~\ref{tab:results}, our model outperforms other methods across all tasks based on PSNR, SSIM, and MSE metrics. Our method achieves PSNR improvements of $1.5 \sim 2.3$ dB and SSIM increases of $2\% \sim 4\%$ in the single-input scenario, surpassing state-of-the-art methods. In the multi-input scenario, the enhanced information further improves performance, with our model achieving a PSNR improvement of $0.9$ dB and an SSIM increase of $2\%$. These results underscore the effectiveness of our approach in cross-modality synthesis tasks.

\begin{table}[ht!]
\centering
\caption{The ablation experiment on the fusion modules, the best results are highlighted in bold.}
\begin{tabular}{cl|lll}
\hline
\multicolumn{2}{c|}{Method}                                         & \multicolumn{1}{c}{PSNR$\uparrow$} & \multicolumn{1}{c}{SSIM$\uparrow$} & \multicolumn{1}{c}{NMSE$\downarrow$} \\ \hline
\multicolumn{1}{c|}{\multirow{3}{*}{T1,T2→T1ce}} & w/o LF           & 29.84$\pm$3.49           & 0.902$\pm$0.042                    & 0.158         \\
\multicolumn{1}{c|}{}                            & w/o GF           & 29.53$\pm$3.35           & 0.900$\pm$0.042                    & 0.181                  \\
\multicolumn{1}{c|}{}                            & Ours             & \textbf{30.22$\pm$3.45}      & \textbf{0.909$\pm$0.040} & \textbf{0.138}                  \\ \hline
\end{tabular}
\label{tab:ablation_fusion}
\end{table}

\subsection{Ablation Studies}

\textbf{Impact of LF and GF:}
We evaluated the effectiveness of the two fusion modules, LF and GF, with quantitative results presented in Table~\ref{tab:ablation_fusion}. Notably, after removing the Local Feature Fusion (LF) module, we used concatenation to merge features from the two modalities. Compared to our full model, the removal of either fusion module led to a performance drop, indicating that the two fusion modules are complementary. Furthermore, the performance degradation caused by the removal of GF was more pronounced, which may be because GF selectively interacts with the features of the two modalities, allowing each modality to focus on relevant information from the other modality. This compensates for the limitation of convolution, which can only perform feature fusion locally.

\begin{table}[ht!]
\centering
\caption{The effect of varying the number of downsampling operations on the BRATS2021 dataset. `Levels' denote the number of downsampling stages. The best results are highlighted in bold.}
\begin{tabular}{c|l|lll}
\hline
Task                        & \multicolumn{1}{l|}{Levels} & \multicolumn{1}{c}{PSNR$\uparrow$} & \multicolumn{1}{c}{SSIM$\uparrow$} & \multicolumn{1}{c}{NMSE$\downarrow$} \\ \hline
\multirow{3}{*}{T1→T1ce}    & 1                                       & 28.33$\pm$3.34           & 0.879$\pm$0.049                    & 0.191                 \\
                            & 2 (ours)                                      & \textbf{28.81$\pm$3.25}      & \textbf{0.885$\pm$0.045} & \textbf{0.185}        \\
                            & 3                                       & 28.46$\pm$3.27                    & 0.882$\pm$0.046                    & 0.189                 \\ \hline
\multirow{3}{*}{T2→T1ce}    & 1                                       & 28.16$\pm$3.44                    & 0.873$\pm$0.052                    & 0.194                 \\
                            & 2 (ours)                                       & \textbf{28.60$\pm$3.40}    & \textbf{0.880$\pm$0.050}& \textbf{0.186}                 \\
                            & 3                                       & 28.41$\pm$3.45                    & 0.877$\pm$0.056                    & 0.188        \\ \hline
\multirow{3}{*}{T1,T2→T1ce} & 1                                       & 29.64$\pm$3.50                    & 0.899$\pm$0.043                    & 0.151                  \\
                            & 2 (ours)                                       & \textbf{30.22$\pm$3.45}      & \textbf{0.909$\pm$0.040}& \textbf{0.138}         \\
                            & 3                                       & 29.89$\pm$3.45                    & 0.903$\pm$0.044                    & 0.142                  \\ \hline
\end{tabular}
\label{tab:ablation_downsample}
\end{table}

\textbf{Model depth:}
We investigated the impact of varying model depth (number of downsampling stages) on model performance, aiming to achieve optimal performance for the TLP model. The number of downsampling stages determines the minimum resolution processed by the model. The experimental results in Table~\ref{tab:ablation_downsample} demonstrate that two downsampling stages yield the best performance for specific resolutions of brain MRI images.

\begin{table}[ht!]
\centering
\caption{The ablation experiment on FPG, the best results are highlighted in bold.}
\begin{tabular}{cl|lll}
\hline
\multicolumn{2}{c|}{Method}                                         & \multicolumn{1}{c}{PSNR$\uparrow$} & \multicolumn{1}{c}{SSIM$\uparrow$} & \multicolumn{1}{c}{NMSE$\downarrow$} \\ \hline
\multicolumn{1}{c|}{\multirow{2}{*}{T1→T1ce}}    & w/o FPG          & 27.94$\pm$3.73                    & 0.872$\pm$0.054                    & 0.231        \\
\multicolumn{1}{c|}{}                            & Ours             & \textbf{28.81$\pm$3.25}      & \textbf{0.885$\pm$0.045} & \textbf{0.185}               \\ \hline
\multicolumn{1}{c|}{\multirow{2}{*}{T2→T1ce}}    & w/o FPG          & 27.96$\pm$3.69                    & 0.874$\pm$0.052                    & 0.214               \\
\multicolumn{1}{c|}{}                            & Ours             & \textbf{28.60$\pm$3.40}    & \textbf{0.880$\pm$0.050}& \textbf{0.186}               \\ \hline
\multicolumn{1}{c|}{\multirow{2}{*}{T1,T2→T1ce}} & w/o FPG          & 29.32$\pm$3.51                    & 0.898$\pm$0.046                    & 0.149               \\
\multicolumn{1}{c|}{}                            & Ours             & \textbf{30.22$\pm$3.45}      & \textbf{0.909$\pm$0.040}& \textbf{0.138}                \\ \hline
\end{tabular}
\label{tab:ablation_prompt}
\end{table}

\textbf{Impact of FPG:}
Here, we evaluate the impact of FPG on model performance. Our baseline comparison method removes the FPG module and directly uses ground-truth segmentation masks as deterministic prompts during training, entirely excluding perturbation strategies. It's important to note that no prompts are used during the testing phase. Quantitative results presented in Table~\ref{tab:ablation_prompt} reveal a substantial performance degradation when FPG is removed. This significant decline stems from the model's over-reliance on deterministic prompt patterns, which limits its ability to extract features from T1w or T2w images effectively. The experimental findings highlight that FPG plays a crucial role in simulating inter-clinician variability through its innovative design. Not only does FPG enhance the model's adaptability to diverse annotation styles, but it also decouples the network's dependence on specific prompt characteristics, preserving diagnostic accuracy. It ensures robust generalization across various clinical scenarios with differing annotation protocols.

\subsection{Interactive Experiment}\label{subsec:Interactive}

\begin{figure}[ht!] 
\centering
\includegraphics[width=0.48\textwidth]{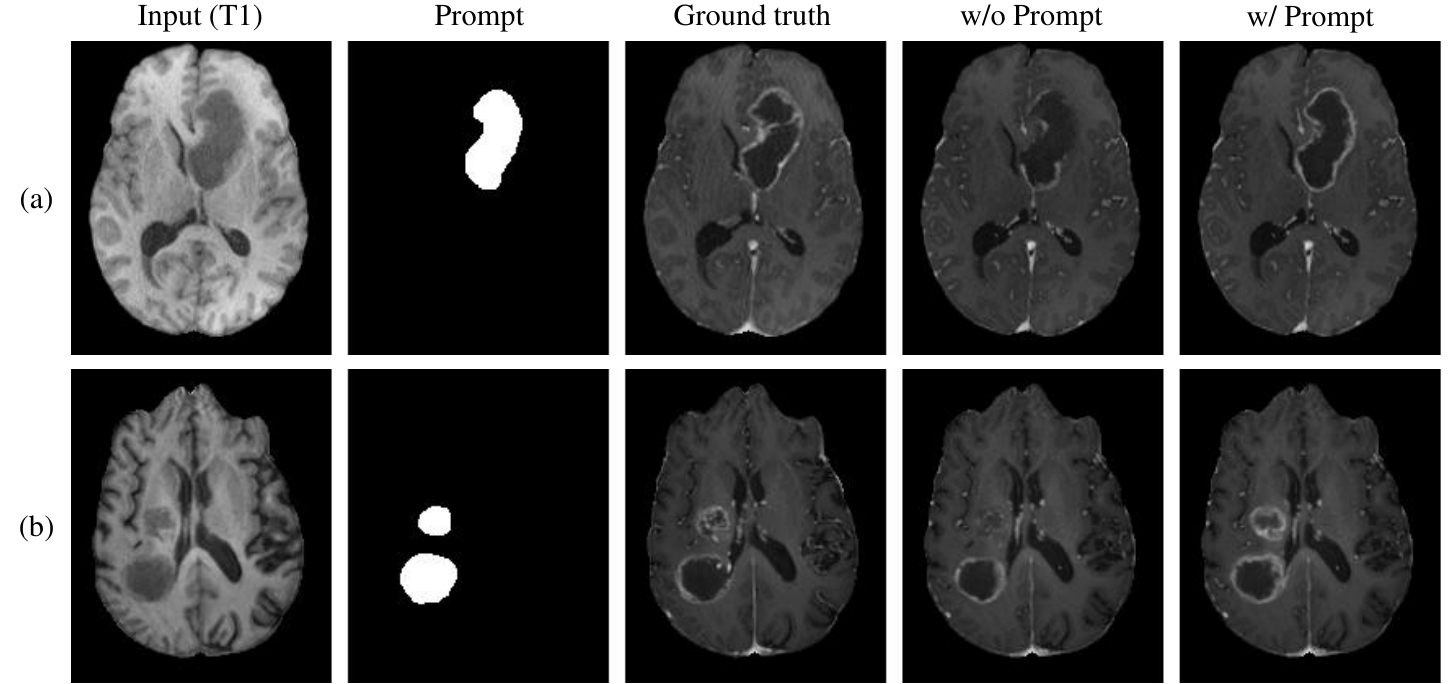}
\caption{The effect of prompts is demonstrated when handling complex tumor shapes.}
\label{fig:prompt}
\end{figure}

Interactivity plays a crucial role in enhancing the performance of our model. As shown in Figure \ref{fig:prompt}, the use of prompt information significantly improves the handling of more complex tumors. In case (a), when no prompts are provided, the model struggles to accurately delineate the complex contours of the tumor. However, when an experienced radiologist manually marks the shadow regions on the T1 image and inputs them as prompts into the network, a notable improvement in the generated results is observed. In case (b), when handling multiple lesions, capturing all of the lesions proves to be challenging.  We suspect that this issue arises from data imbalance, as cases with multiple lesions are relatively rare. In this situation, radiologists can mark all suspicious areas to assist the generation process. The newly generated results, in turn, successfully capture all lesions without any omissions.

\begin{figure}[ht!] 
\centering
\includegraphics[width=0.48\textwidth]{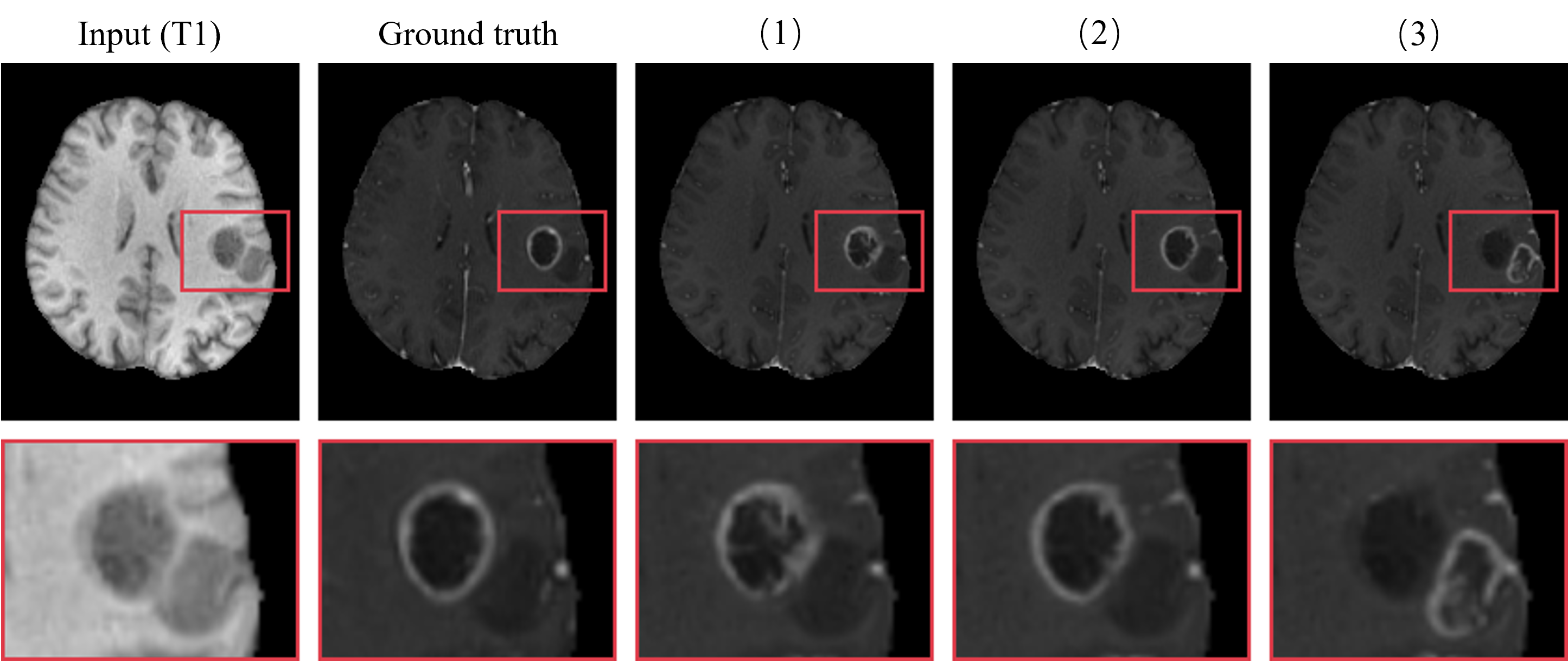}
\caption{Changing the prompt information can produce many interesting results. (1) is the image generated without any prompt, (2) is the image generated with the correct prompt, and (3) is the image generated with an intentionally incorrect prompt.}
\label{fig:Interactive}
\end{figure}

Additionally, an interactive model opens up intriguing research opportunities. In the case of the two elliptical shadows shown in Fig. \ref{fig:Interactive}, individuals without professional expertise might mistakenly identify both as tumors, when in fact only the inner shadow represents a tumor. Despite this, our model accurately enhance the tumor's location without any prompts (see Fig.\ref{fig:Interactive}-(1)). When the correct prompt is provided by circling the inner shadow, the generated result becomes significantly clearer (see Fig.~\ref{fig:Interactive}-(2)). However, when another shadow region is circled as the prompt, the model produces an unclear, erroneous result (see Fig.\ref{fig:Interactive}-(3), the quality of this erroneous result is clearly insufficient. This demonstrates that TLP offers better interpretability, enhancing our understanding of specific cases. Furthermore, the model's ability to generate diverse results underscores TLP's significant advantage in data augmentation for downstream tasks.
\section{Discussion}
The development of interactive models represents a crucial research direction in medical imaging, with our proposed framework demonstrating significant value through multiple critical dimensions:

\textbf{Human-AI Knowledge Integration:} Current deep generative models in medicine exhibit fundamental limitations in human-AI interaction. Conventional models produce non-modifiable outputs post-training, even when generating anatomically implausible results, leaving radiologists without mechanisms to adjust suspicious findings. Our interactive paradigm addresses this critical gap by enabling radiologists to focus on specific ROIs,  particularly valuable for low-quality inputs where diagnostic uncertainty increases. This capability not only enhances diagnostic workflows but also facilitates comprehensive lesion exploration, as we extensively demonstrate through various interesting results in Section \ref{subsec:Interactive}.

\textbf{Downstream Task Enhancement:} Medical image generation achieves its full potential when effectively serving subsequent clinical tasks including tumor classification, segmentation, and detection. While traditional generative approaches offer limited data augmentation capabilities through deterministic outputs, our interactive model enables order-of-magnitude improvements in dataset diversity through prompt engineering. By modifying the ROI prompts, radiologists can obtain targeted variations that address specific data deficiencies in downstream models, moving beyond simple dataset expansion to intentional quality enhancement.

\textbf{Training Robustness:} Our framework avoids the inherent key stability issues in the joint training approach with segmentation models \cite{yang2023learning}. Unlike methods susceptible to error propagation from imperfect segmentation outputs, our design intentionally decouples generation guidance from automated segmentation results. This architecture acknowledges the variability in clinical annotations—differences in tumor boundaries often exist among radiologists and across various annotation platforms. Through our FPG, we simulate this natural variation while preventing model overreliance on any single annotation pattern, thereby improving generalization capabilities.

\textbf{Resource Efficiency:} The implementation of interactive systems traditionally faces substantial data annotation challenges in medical domains. Our innovative FPG framework significantly reduces dependency on expensive expert annotations through algorithmically generated prompts while maintaining clinical relevance. Through multi-directional convolutional operations that mimic radiologists' annotation patterns (expansion/contraction, focus shifting), we achieve cost-effective training without compromising output quality. We note that optimal performance would still require expert-validated annotations, suggesting promising future research directions in hybrid human-AI annotation systems.

The FPG's design philosophy emphasizes clinical adaptability through simulated annotation variability. While our current implementation uses basic spatial transformations, future improvements could incorporate more sophisticated annotation patterns learned from actual radiologists' behavior. Developing attention-aware algorithms that prioritize clinically relevant features during prompt generation could further bridge the gap between computational efficiency and clinical utility, potentially establishing new standards for interactive medical imaging systems.

\section{Conclusion}
In this study, we propose the TLP framework for CE-MRI synthesis, to reduce or eliminate the need for GBCAs. The Fuzzy Prompt Generation (FPG) module is introduced to generate prompt information that aids in the training process. Additionally, an efficient Transformer-based backbone and two novel fusion modules are designed to make significant strides in handling tumors of varying sizes while preserving critical anatomical details. To the best of our knowledge, TLP represents the first interactive CE-MRI synthesis paradigm, allowing radiologists to refine outputs through targeted anatomical prompts, thereby integrating human prior knowledge. This interactivity opens up exciting new applications, which we validate in Section \ref{subsec:Interactive}. This innovation bridges a critical gap in clinical AI systems, offering a feasible solution that addresses the toxicity concerns associated with GBCAs while maintaining the practicalities of the diagnostic workflow by combining diagnostic reliability with physician-guided adaptability.
\bibliographystyle{ieeetr}
\bibliography{references}

\begin{thebibliography}{10}

\bibitem{grobner2006gadolinium}
T.~Grobner, ``Gadolinium--a specific trigger for the development of nephrogenic fibrosing dermopathy and nephrogenic systemic fibrosis?,'' {\em Nephrology Dialysis Transplantation}, vol.~21, no.~4, pp.~1104--1108, 2006.

\bibitem{warntjes2018synthesizing}
M.~Warntjes, I.~Blystad, A.~Tisell, and E.-M. Larsson, ``Synthesizing a contrast-enhancement map in patients with high-grade gliomas based on a postcontrast mr imaging quantification only,'' {\em American Journal of Neuroradiology}, vol.~39, no.~12, pp.~2194--2199, 2018.

\bibitem{zahra2007dynamic}
M.~A. Zahra, K.~G. Hollingsworth, E.~Sala, D.~J. Lomas, and L.~T. Tan, ``Dynamic contrast-enhanced mri as a predictor of tumour response to radiotherapy,'' {\em The lancet oncology}, vol.~8, no.~1, pp.~63--74, 2007.

\bibitem{silver1997sensitivity}
N.~Silver, C.~Good, G.~Barker, D.~MacManus, A.~Thompson, I.~Moseley, W.~McDonald, and D.~Miller, ``Sensitivity of contrast enhanced mri in multiple sclerosis. effects of gadolinium dose, magnetization transfer contrast and delayed imaging.,'' {\em Brain: a journal of neurology}, vol.~120, no.~7, pp.~1149--1161, 1997.

\bibitem{khan2014molecular}
U.~A. Khan, L.~Liu, F.~A. Provenzano, D.~E. Berman, C.~P. Profaci, R.~Sloan, R.~Mayeux, K.~E. Duff, and S.~A. Small, ``Molecular drivers and cortical spread of lateral entorhinal cortex dysfunction in preclinical alzheimer's disease,'' {\em Nature neuroscience}, vol.~17, no.~2, pp.~304--311, 2014.

\bibitem{moya2024exogenous}
E.~Moya-S{\'a}ez, R.~de~Luis-Garc{\'\i}a, L.~Nunez-Gonzalez, C.~Alberola-L{\'o}pez, and J.~A. Hern{\'a}ndez-Tamames, ``Exogenous agent-free synthetic post-contrast imaging with a cascade of deep networks for enhancement prediction after tumor resection. a parametric-map oriented approach,'' in {\em International Workshop on Simulation and Synthesis in Medical Imaging}, pp.~113--123, Springer, 2024.

\bibitem{thomsen2006nephrogenic}
H.~S. Thomsen, ``Nephrogenic systemic fibrosis: a serious late adverse reaction to gadodiamide,'' {\em European radiology}, vol.~16, no.~12, pp.~2619--2621, 2006.

\bibitem{marckmann2006nephrogenic}
P.~Marckmann, L.~Skov, K.~Rossen, A.~Dupont, M.~B. Damholt, J.~G. Heaf, and H.~S. Thomsen, ``Nephrogenic systemic fibrosis: suspected causative role of gadodiamide used for contrast-enhanced magnetic resonance imaging,'' {\em Journal of the American Society of Nephrology}, vol.~17, no.~9, pp.~2359--2362, 2006.

\bibitem{forghani2016adverse}
R.~Forghani, ``Adverse effects of gadolinium-based contrast agents: changes in practice patterns,'' {\em Topics in Magnetic Resonance Imaging}, vol.~25, no.~4, pp.~163--169, 2016.

\bibitem{olchowy2017presence}
C.~Olchowy, K.~Cebulski, M.~{\L}asecki, R.~Chaber, A.~Olchowy, K.~Ka{\l}wak, and U.~Zaleska-Dorobisz, ``The presence of the gadolinium-based contrast agent depositions in the brain and symptoms of gadolinium neurotoxicity-a systematic review,'' {\em PloS one}, vol.~12, no.~2, p.~e0171704, 2017.

\bibitem{semelka2016gadolinium}
R.~C. Semelka, J.~Ramalho, A.~Vakharia, M.~AlObaidy, L.~M. Burke, M.~Jay, and M.~Ramalho, ``Gadolinium deposition disease: initial description of a disease that has been around for a while,'' {\em Magnetic resonance imaging}, vol.~34, no.~10, pp.~1383--1390, 2016.

\bibitem{purysko2011focal}
A.~Purysko, E.~Remer, and J.~Veniero, ``Focal liver lesion detection and characterization with gd-eob-dtpa,'' {\em Clinical radiology}, vol.~66, no.~7, pp.~673--684, 2011.

\bibitem{zhang2024unified}
Y.~Zhang, C.~Peng, Q.~Wang, D.~Song, K.~Li, and S.~K. Zhou, ``Unified multi-modal image synthesis for missing modality imputation,'' {\em IEEE Transactions on Medical Imaging}, 2024.

\bibitem{jiao2023contrast}
C.~Jiao, D.~Ling, S.~Bian, A.~Vassantachart, K.~Cheng, S.~Mehta, D.~Lock, Z.~Zhu, M.~Feng, H.~Thomas, {\em et~al.}, ``Contrast-enhanced liver magnetic resonance image synthesis using gradient regularized multi-modal multi-discrimination sparse attention fusion gan,'' {\em Cancers}, vol.~15, no.~14, p.~3544, 2023.

\bibitem{ye2013modality}
D.~H. Ye, D.~Zikic, B.~Glocker, A.~Criminisi, and E.~Konukoglu, ``Modality propagation: coherent synthesis of subject-specific scans with data-driven regularization,'' in {\em Medical Image Computing and Computer-Assisted Intervention--MICCAI 2013: 16th International Conference, Nagoya, Japan, September 22-26, 2013, Proceedings, Part I 16}, pp.~606--613, Springer, 2013.

\bibitem{huang2017simultaneous}
Y.~Huang, L.~Shao, and A.~F. Frangi, ``Simultaneous super-resolution and cross-modality synthesis of 3d medical images using weakly-supervised joint convolutional sparse coding,'' in {\em Proceedings of the IEEE conference on computer vision and pattern recognition}, pp.~6070--6079, 2017.

\bibitem{van2015cross}
H.~Van~Nguyen, K.~Zhou, and R.~Vemulapalli, ``Cross-domain synthesis of medical images using efficient location-sensitive deep network,'' in {\em Medical Image Computing and Computer-Assisted Intervention--MICCAI 2015: 18th International Conference, Munich, Germany, October 5-9, 2015, Proceedings, Part I 18}, pp.~677--684, Springer, 2015.

\bibitem{zhang2023synthesis}
T.~Zhang, L.~Han, A.~D’Angelo, X.~Wang, Y.~Gao, C.~Lu, J.~Teuwen, R.~Beets-Tan, T.~Tan, and R.~Mann, ``Synthesis of contrast-enhanced breast mri using t1-and multi-b-value dwi-based hierarchical fusion network with attention mechanism,'' in {\em International Conference on Medical Image Computing and Computer-Assisted Intervention}, pp.~79--88, Springer, 2023.

\bibitem{zhao2020tripartite}
J.~Zhao, D.~Li, Z.~Kassam, J.~Howey, J.~Chong, B.~Chen, and S.~Li, ``Tripartite-gan: Synthesizing liver contrast-enhanced mri to improve tumor detection,'' {\em Medical image analysis}, vol.~63, p.~101667, 2020.

\bibitem{li2022virtual}
W.~Li, H.~Xiao, T.~Li, G.~Ren, S.~Lam, X.~Teng, C.~Liu, J.~Zhang, F.~K.-h. Lee, K.-h. Au, {\em et~al.}, ``Virtual contrast-enhanced magnetic resonance images synthesis for patients with nasopharyngeal carcinoma using multimodality-guided synergistic neural network,'' {\em International Journal of Radiation Oncology* Biology* Physics}, vol.~112, no.~4, pp.~1033--1044, 2022.

\bibitem{sevetlidis2016whole}
V.~Sevetlidis, M.~V. Giuffrida, and S.~A. Tsaftaris, ``Whole image synthesis using a deep encoder-decoder network,'' in {\em Simulation and Synthesis in Medical Imaging: First International Workshop, SASHIMI 2016, Held in Conjunction with MICCAI 2016, Athens, Greece, October 21, 2016, Proceedings 1}, pp.~127--137, Springer, 2016.

\bibitem{chartsias2017multimodal}
A.~Chartsias, T.~Joyce, M.~V. Giuffrida, and S.~A. Tsaftaris, ``Multimodal mr synthesis via modality-invariant latent representation,'' {\em IEEE transactions on medical imaging}, vol.~37, no.~3, pp.~803--814, 2017.

\bibitem{joyce2017robust}
T.~Joyce, A.~Chartsias, and S.~A. Tsaftaris, ``Robust multi-modal mr image synthesis,'' in {\em Medical Image Computing and Computer Assisted Intervention- MICCAI 2017: 20th International Conference, Quebec City, QC, Canada, September 11-13, 2017, Proceedings, Part III 20}, pp.~347--355, Springer, 2017.

\bibitem{ji2022synthetic}
S.~Ji, D.~Yang, J.~Lee, S.~H. Choi, H.~Kim, and K.~M. Kang, ``Synthetic mri: technologies and applications in neuroradiology,'' {\em Journal of Magnetic Resonance Imaging}, vol.~55, no.~4, pp.~1013--1025, 2022.

\bibitem{preetha2021deep}
C.~J. Preetha, H.~Meredig, G.~Brugnara, M.~A. Mahmutoglu, M.~Foltyn, F.~Isensee, T.~Kessler, I.~Pfl{\"u}ger, M.~Schell, U.~Neuberger, {\em et~al.}, ``Deep-learning-based synthesis of post-contrast t1-weighted mri for tumour response assessment in neuro-oncology: a multicentre, retrospective cohort study,'' {\em The Lancet Digital Health}, vol.~3, no.~12, pp.~e784--e794, 2021.

\bibitem{isola2017image}
P.~Isola, J.-Y. Zhu, T.~Zhou, and A.~A. Efros, ``Image-to-image translation with conditional adversarial networks,'' in {\em Proceedings of the IEEE conference on computer vision and pattern recognition}, pp.~1125--1134, 2017.

\bibitem{zhu2017unpaired}
J.-Y. Zhu, T.~Park, P.~Isola, and A.~A. Efros, ``Unpaired image-to-image translation using cycle-consistent adversarial networks,'' in {\em Proceedings of the IEEE international conference on computer vision}, pp.~2223--2232, 2017.

\bibitem{zhou2020hi}
T.~Zhou, H.~Fu, G.~Chen, J.~Shen, and L.~Shao, ``Hi-net: hybrid-fusion network for multi-modal mr image synthesis,'' {\em IEEE transactions on medical imaging}, vol.~39, no.~9, pp.~2772--2781, 2020.

\bibitem{wang2018non}
X.~Wang, R.~Girshick, A.~Gupta, and K.~He, ``Non-local neural networks,'' in {\em Proceedings of the IEEE conference on computer vision and pattern recognition}, pp.~7794--7803, 2018.

\bibitem{kodali2017convergence}
N.~Kodali, J.~Abernethy, J.~Hays, and Z.~Kira, ``On convergence and stability of gans,'' {\em arXiv preprint arXiv:1705.07215}, 2017.

\bibitem{atli2024i2i}
O.~F. Atli, B.~Kabas, F.~Arslan, M.~Yurt, O.~Dalmaz, and T.~{\c{C}}ukur, ``I2i-mamba: Multi-modal medical image synthesis via selective state space modeling,'' {\em arXiv preprint arXiv:2405.14022}, 2024.

\bibitem{xie2021cotr}
Y.~Xie, J.~Zhang, C.~Shen, and Y.~Xia, ``Cotr: Efficiently bridging cnn and transformer for 3d medical image segmentation,'' in {\em Medical Image Computing and Computer Assisted Intervention--MICCAI 2021: 24th International Conference, Strasbourg, France, September 27--October 1, 2021, Proceedings, Part III 24}, pp.~171--180, Springer, 2021.

\bibitem{chen2021transunet}
J.~Chen, Y.~Lu, Q.~Yu, X.~Luo, E.~Adeli, Y.~Wang, L.~Lu, A.~L. Yuille, and Y.~Zhou, ``Transunet: Transformers make strong encoders for medical image segmentation,'' {\em arXiv preprint arXiv:2102.04306}, 2021.

\bibitem{dalmaz2022resvit}
O.~Dalmaz, M.~Yurt, and T.~{\c{C}}ukur, ``Resvit: residual vision transformers for multimodal medical image synthesis,'' {\em IEEE Transactions on Medical Imaging}, vol.~41, no.~10, pp.~2598--2614, 2022.

\bibitem{liu2021swin}
Z.~Liu, Y.~Lin, Y.~Cao, H.~Hu, Y.~Wei, Z.~Zhang, S.~Lin, and B.~Guo, ``Swin transformer: Hierarchical vision transformer using shifted windows,'' in {\em Proceedings of the IEEE/CVF international conference on computer vision}, pp.~10012--10022, 2021.

\bibitem{zamir2022restormer}
S.~W. Zamir, A.~Arora, S.~Khan, M.~Hayat, F.~S. Khan, and M.-H. Yang, ``Restormer: Efficient transformer for high-resolution image restoration,'' in {\em Proceedings of the IEEE/CVF conference on computer vision and pattern recognition}, pp.~5728--5739, 2022.

\bibitem{ozbey2023unsupervised}
M.~{\"O}zbey, O.~Dalmaz, S.~U. Dar, H.~A. Bedel, {\c{S}}.~{\"O}zturk, A.~G{\"u}ng{\"o}r, and T.~{\c{C}}ukur, ``Unsupervised medical image translation with adversarial diffusion models,'' {\em IEEE Transactions on Medical Imaging}, 2023.

\bibitem{jiang2024fast}
H.~Jiang, M.~Imran, L.~Ma, T.~Zhang, Y.~Zhou, M.~Liang, K.~Gong, and W.~Shao, ``Fast-ddpm: Fast denoising diffusion probabilistic models for medical image-to-image generation,'' {\em arXiv e-prints}, pp.~arXiv--2405, 2024.

\bibitem{xu2021synthesis}
C.~Xu, D.~Zhang, J.~Chong, B.~Chen, and S.~Li, ``Synthesis of gadolinium-enhanced liver tumors on nonenhanced liver mr images using pixel-level graph reinforcement learning,'' {\em Medical image analysis}, vol.~69, p.~101976, 2021.

\bibitem{radford2021learning}
A.~Radford, J.~W. Kim, C.~Hallacy, A.~Ramesh, G.~Goh, S.~Agarwal, G.~Sastry, A.~Askell, P.~Mishkin, J.~Clark, {\em et~al.}, ``Learning transferable visual models from natural language supervision,'' in {\em International conference on machine learning}, pp.~8748--8763, PMLR, 2021.

\bibitem{rombach2022high}
R.~Rombach, A.~Blattmann, D.~Lorenz, P.~Esser, and B.~Ommer, ``High-resolution image synthesis with latent diffusion models,'' in {\em Proceedings of the IEEE/CVF conference on computer vision and pattern recognition}, pp.~10684--10695, 2022.

\bibitem{ramesh2022hierarchical}
A.~Ramesh, P.~Dhariwal, A.~Nichol, C.~Chu, and M.~Chen, ``Hierarchical text-conditional image generation with clip latents,'' {\em arXiv preprint arXiv:2204.06125}, vol.~1, no.~2, p.~3, 2022.

\bibitem{saharia2022photorealistic}
C.~Saharia, W.~Chan, S.~Saxena, L.~Li, J.~Whang, E.~L. Denton, K.~Ghasemipour, R.~Gontijo~Lopes, B.~Karagol~Ayan, T.~Salimans, {\em et~al.}, ``Photorealistic text-to-image diffusion models with deep language understanding,'' {\em Advances in neural information processing systems}, vol.~35, pp.~36479--36494, 2022.

\bibitem{park2019semantic}
T.~Park, M.-Y. Liu, T.-C. Wang, and J.-Y. Zhu, ``Semantic image synthesis with spatially-adaptive normalization,'' in {\em Proceedings of the IEEE/CVF conference on computer vision and pattern recognition}, pp.~2337--2346, 2019.

\bibitem{nichol2021glide}
A.~Nichol, P.~Dhariwal, A.~Ramesh, P.~Shyam, P.~Mishkin, B.~McGrew, I.~Sutskever, and M.~Chen, ``Glide: Towards photorealistic image generation and editing with text-guided diffusion models,'' {\em arXiv preprint arXiv:2112.10741}, 2021.

\bibitem{kingma2014adam}
D.~P. Kingma, ``Adam: A method for stochastic optimization,'' {\em arXiv preprint arXiv:1412.6980}, 2014.

\bibitem{wang2004image}
Z.~Wang, A.~C. Bovik, H.~R. Sheikh, and E.~P. Simoncelli, ``Image quality assessment: from error visibility to structural similarity,'' {\em IEEE transactions on image processing}, vol.~13, no.~4, pp.~600--612, 2004.

\bibitem{yang2023learning}
H.~Yang, J.~Sun, and Z.~Xu, ``Learning unified hyper-network for multi-modal mr image synthesis and tumor segmentation with missing modalities,'' {\em IEEE Transactions on Medical Imaging}, 2023.

\end{thebibliography}

\end{document}